\documentstyle[12pt]{article}
\begin{document}
\renewcommand{\theequation}{\arabic{section}.\arabic{equation}}
\setcounter{equation}{0}


\def\wh{\widehat}
\def\wt{\widetilde}
\def\D{\Cal{D}}
\def\ov{\overline}
\def\un{\underline}
\def\noi{\noindent}


\centerline{\huge On the Stability of Compactified}
\centerline{\huge D=11 Supermembranes}

\vskip 2.5cm

\centerline{I. Mart\'{\i}n}

\centerline{\it Universidad Sim\'{o}n Bol\'{\i}var, Departamento de F\'{\i}sica}

\centerline{\it Caracas 89000, Venezuela.}
\centerline{\it e-mail:isbeliam@usb.ve}

\vskip 1cm

\centerline{A. Restuccia}

\centerline{\it Universidad Sim\'{o}n Bol\'{\i}var, Departamento de F\'{\i}sica}

\centerline{\it Caracas 89000, Venezuela.}
\centerline{\it e-mail: arestu@usb.ve}

\vskip 1cm

\centerline{R. Torrealba}

\centerline{\it Depto. de Matem\'{a}ticas , Decanato de Ciencias,}

\centerline{\it Universidad Centro Occidental "Lisandro Alvarado",
Apartado 400,}

\centerline{\it Barquisimeto, Venezuela.}
\centerline{\it e-mail:rtorre@compaq.ucla.edu.ve}

\vskip .6cm
\begin{abstract}
We prove  $D=11$ supermembrane theories wrapping around in an 
irreducible way over  $S^{1} \times S^{1}\times M^{9}$ on the target
manifold, have a hamiltonian with strict  minima and without infinite 
dimensional valleys at the minima for the bosonic sector. The minima occur at monopole 
connections of an associated $U(1)$ bundle over topologically non trivial
Riemann surfaces of arbitrary genus. Explicit expressions for the minimal
connections in terms of membrane maps are presented. The minimal maps and 
corresponding connections satisfy the BPS condition with half SUSY.
\end{abstract}
\centerline{\bf June 1997}

\newpage
\section{Introduction}

Minkowski $D=11$ supermembrane theories are thought to be unstable at the quantum
level because of the presence of string like configurations that, with  
supersymmetry, renders the spectrum continuous \cite{N + de Wit}.There the problem
of the stability of quantum supermembranes was addressed by approximating the
supermembrane action by a $SU(N)$ super Yang Mills theory as $N\rightarrow 
\infty $. Even when their result looks plausible, the whole proof depends on
whether the $N\rightarrow \infty $ is applicable. From the point of view 
of D=brane theories \cite{Pol}, however, what it should be relevant is not 
the supermembrane with 11-dimensional Minkowskian space as target space 
but a compactified version, where at least one dimension is compactified 
to $S^{1}$.   
In  recent works by  \cite{de Wit} \cite{Russo}
the $SU(N)$ super Yang Mills theory approach to the compactified supermembrane
is taken but without a clear conclusion on the problem of stability. In 
\cite{Russo} it is argued that winding over more than one compact 
dimension may remove the continuity of the spectrum. In \cite{de Wit} 
however the opposite argument is given based on the qualitative feature 
that classical membranes with winding can still have string like 
configurations of arbitrary length without increasing its area. But the 
problem remains with no solution since a generalization of the $SU(N)$
supersymmetric matrix model regularization to the compactified version was not
found.

In this paper, we analyze the bosonic part of the Hamiltonian of the 
compactified supermembrane from a different point of view. Tackling the problem
geometrically in terms of $U(1)$ connections over non trivial bundles 
defined on Riemann surfaces of arbitrary genus $g$.

 We make use of an irreducibility condition on the winding of the
supermembrane on the compactified directions in the target 
space. When this condition is not satisfied it is straightforward to show 
the existence of infinite dimensional valleys in the bosonic sector of the 
supermembrane Hamiltonian leading to a continuous spectrum as argued in
 \cite{de Wit}.
 
  We prove that area preserving diffeomorphisms on
the supermembrane may be lifted to gauge transformations of particular $U(1)$
bundles .So that the problem of minima in the mass operator of the supermembrane
may be addressed from the point of view of an equivalent Maxwell action. We
conclude that the minima space is finite dimensional, with no minimal valleys that go 
to infinity, at least for the bosonic contribution of the Hamiltonian. Also, we
construct the explicit solutions of these minima corresponding to filaments of
 $U(1)$ monopoles.
 
 In section 2, we briefly present the problem of the existence of infinite 
 dimensional valleys at the minimum of the Hamiltonian. In section 3, we 
 formulate the problem in term of connections constructed from membrane 
 maps and introduce the irreducibility condition. Next, we explicitly 
 construct the minimal connections in terms of monopole potentials. It is 
 shown that they provide a minima of the Hamiltonian. In section 5, we 
 show that the space of minimal maps reconstructed from minimal connections 
 is finite dimensional.

\setcounter{equation}{0}\section{The unstable Supermembrane}
\noindent
 
The hamiltonian of the $D=11$ Supermembrane, in the light cone gauge, 
takes the form

\begin{equation}
	H=\frac{P_{oa}^{2}}{2 P_{o}^{+}}+ \frac{M^{2}}{2P_{o}^{+}}
\end{equation}
where $P_{0a}$ are the transverse momenta of the center of mass and $M$ 
is the Supermembrane mass operator.

The bosonic contribution to the mass operator is

\begin{eqnarray}
	{\cal M}^{2}& = & \int_{\Sigma}d^{2}\sigma \sqrt{g}\,[P_{a}^{2}+ 
	V(x)],\nonumber \\
	V(x) & = & \frac{1}{2}(\epsilon^{rj} 
	\partial_{r}X_{a}\partial_{j}X_{b})^{2}
\end{eqnarray}
where $\sigma^{r}$, r=1,2, denote local coordinates on the 2-dimensional 
world volume $\Sigma$.$ X^{a}, P_{a} a=1,\ldots ,9$ are the transverse 
coordinates and conjugate momenta of the supermembrane in the light 
cone coordinates.

The world volume $\Sigma $ is usually taken to be a Riemann surface of genus g.
The analysis in \cite{N + de Wit} is based on an expansion of coordinates 
and momenta in terms of a complete set of real orthonormal functions 
$Y^{A}(\sigma), A=1,\ldots, \infty $. This set represent the generators of 
a $SU(N)$ gauge theory in the limit when $N$ tends to infinity. In 
particular, the coordinates are expressed as 

\begin{equation}
X^{a}(\sigma)= \sum_{A}	X_{A}^{a}Y^{A}(\sigma)
\end{equation}
The mass operator then takes the form
\begin{equation}
M^{^2}= (P_{a}^{A})+(F_{ABC}X_{a}^{B}X_{b}^{B})^{2}- if_{ABC}\theta 
^{A} \gamma ^{a} X_{a}^{B} \theta^{C}	
\end{equation}
where $\theta ^{A} $ are the fermionic coordinates, $f_{ABC}$ are the structure constants of the $SU(N)$ gauge theory that 
approximates, in the limit $N \rightarrow \infty$, to the invariant subgroup 
of the area preserving transformations of the supermembrane.
In fact, the group of area preserving diffeomorphisms for spherical membranes and for toroidal membranes 
can be regarded as the limit of $SU(N)$, $N \rightarrow \infty$.  The 
approach of \cite{N + de Wit} is to regularize the theory by replacing 
this infinite dimensional group by $SU(N)$.

An important property of (2.2) is that it vanishes if the coordinates 
$X_{A}^{a}$ take values in some abelian subalgebra.
The classical theory is then unstable  but quantum mechanical 
effects turn  it stable.  This is in fact the case 
for the bosonic membrane.  It is classically  unstable because the 
potential has zero valleys along $X_{a}=0$ for all but one coordinate that 
may have an arbitrary value.  However, the quantum mechanical oscillators
perpendicular to the valley directions give rise to a zero point energy, 
inducing an effective potential barrier which confines the wave function.  
When supersymmetry is introduced,the situation  changes because SUSY harmonic
oscillators have no zero point energy so that the confining potential 
vanishes.  Explicit calculations \cite{N + de Wit} show indeed that the wave 
function can go to infinity along the valleys of zero classical 
energy, so that the spectrum of the supermembrane is continuous. 

 The main ingredients leading to an  unstable problem are then the 
existence of valleys with zero energy extending to 
infinity for any arbitrary map $X(\sigma^{1}, \sigma ^{2})$ with all 
other coordinates $X^{a}$ zero in (2.2), and SUSY which eliminates 
the zero point energy of the bosonic oscillators perpendicular to the valley 
directions.
In the next section we will consider the problem of stability for 
compactified supermembranes.

\setcounter{equation}{0}\section{Minima of the Potential}

On what follows we will address the problem of instability from another 
point of view, taking advantage of the physics of $U(1)$ bundles.  We first 
consider the bosonic contribution to the supermembrane potential.
To analyze the minima of the potential (2.2), we rewrite it putting the momenta equal to 
zero since  it will be satisfied by all minimal configurations, in the following way 

\begin{equation}
	<V(x)>_{\sigma} \equiv \int_{\Sigma}d^{2}\sigma\sqrt{g}\, V(x)= \int_{\sigma} 
{^\ast F_{ab} F_{ab}}
\end{equation}
where
\begin{equation}
F_{ab}\equiv d(X_{a}dX_{b})  ,\, \, a,b = 1,\ldots,9.
\end{equation}
$d$ being the exterior derivative on the world volume $\sigma$, which is 
taken to be a Riemann surface of arbitrary genus g. 
$*F_{ab}$ denotes the Hodge dual of the 2-form $F_{ab}$. The latter may be 
regarded as different `Maxwell fields' for every pair of numbers $ab$.

We assume the supermembrane wraps up in an irreducible way over the torus $S^{1} 
\times S^{1}$ in the target space.  Let $X(\sigma)$ and $Y(\sigma)$ denote the 
compactified coordinates on the target manifold. $X(\sigma)$ defines a map 
from the world volume $\Sigma$ to $S^{1}$. $dX$ may then always be expressed 
as
\begin{equation}
dX= -i g^{-1}dg	
\end{equation}
where
\begin{equation}
g= \exp (i\phi)	
\end{equation}
$\varphi$ being an angular coordinate over $S^{1}$.
$dX$ satisfies the following properties for a 1-form L,

\begin{eqnarray}
dL & = &0 \\
\oint_{{\cal C}_{i}}L & = & 2\pi n^{i}
\end{eqnarray}
where $C_{i}$ denotes a basis of the integer homology of dimension 
one over the worldvolume.
The converse is also valid \cite{C+M+R}, given a globally defined 1-form L 
over $\Sigma$ satisfying (3.5) and (3.6) there exists a map $X : 
\Sigma \rightarrow S^{1}$ for which $L= dX$.

We will say that the supermembrane wraps up in a {\underline 
{nontrivial}} way when at least one of the $n^{i}$ is different from 
zero.

We also say that the supermembrane wraps up in an {\underline 
{irreducible}} way over $S^{1} \times S^{1}$ when

\begin{equation}
\frac{1}{2\pi}\int_{\Sigma}	dX \wedge dY \neq 0
\end{equation}

If the supermembrane wraps up in an irreducible way it does it 
in a nontrivial way over each of the $S^{1}$ in $S^{1} \times 
S^{1}$. Moreover,

\begin{equation}
\frac{1}{2\pi}\int_{\Sigma}	dX \wedge dY =2\pi N	
\end{equation}
where N is also an integer obtained as the product of the $n^{i}$, 
associated to conjugate pairs on a canonical basis over $\Sigma$.

We look now for the stationary points of (3.1) over the space of maps 
defining supermembranes with irreducible wrapping over 
$S^{1} \times S^{1}$. It is straightforward to see in this case  that the 
minimal configurations occur when all but $X,Y$ maps are zero. Associated to this space we may introduce an 
$U(1)$ principle bundle.  We proceed by noting that

\begin{equation}
F=\frac{1}{2\pi}dX \wedge dY
\end{equation}
are closed 2-forms globally defined over $\Sigma$ satisfying (3.8).
By Weil's Theorem \cite{H}, \cite{C+M+R}, there exists a $U(1)$ principle
bundle  and a connection over it such that its pull back by sections over 
$\Sigma$ are 1-form connections with curvatures given by (3.9).
The stationary points of (3.1) with respect to the space of 1-form 
connections satisfy

\begin{equation}
d {^\ast F} = 0.	
\end{equation}
now, since $*F$ is a 0-form,

\begin{equation}
^\ast F	= constant.
\end{equation}
using (3.8) for the compactified directions, we get
\begin{equation}
^\ast F =\frac{2\pi N}{Vol\Sigma}=N.	
\end{equation}

Although there may be additionally stationary points when taking 
variations with respect to the maps, the only relevant ones are determined 
by (3.12) . In fact, we will now show that these are the only strict minima of 
the potential (3.1).

Let $A_{0}$ be a 1-form connection satisfying (3.12) and  $A_{1}$ 
another 1-form connection on the same  $U_{1}$ principle 
bundle characterized by N, then the curvature of any connection $A$ on it 
must satisfy
\begin{eqnarray}
	dF & = & 0 \\
	\int_{\Sigma}F & = & 2\pi N
\end{eqnarray}

Consider now that the following connection is also a 1-form connection
on the same principle bundle  

\begin{equation}
(1-\lambda)A_{0} + \lambda A_{1},	
\end{equation}
with $\lambda$ an arbitrary real parameter .

We then have

\begin{equation}
\int_{\Sigma}{ ^\ast F_{\lambda}} F_{\lambda}=(1-{\lambda}^{2})2\pi 
N^{2} + {\lambda}^{2}\int_{\Sigma}{^\ast F(A_{1}) F(A_{1})},
\end{equation}
where $F_{\lambda}$ is the curvature of (3.15).

From (3.16) we obtain

\begin{equation}
\int_{\Sigma} {^\ast F_{\lambda}F_{\lambda}}= 2\pi N^{2} 
+{\lambda}^{2}  \int_{\Sigma}[^\ast F(A_{1}) F(A_{1}) - ^\ast F(A_{0}) F(A_{0})],
\end{equation}
since the left hand member is positive and (3.17) is valid for all 
$\lambda$ we get

\begin{equation}
\int_{\Sigma} {^\ast} F(A_{1}) F(A_{1}) \geq \int_{\Sigma}	^\ast F(A_{0}) F(A_{0})
\end{equation}

Moreover, the equality is obtained if and only if

\begin{equation}
A_{1}=A_{0}+d\Lambda
\end{equation}
where $\Lambda$ is a closed 1-form.

The stationary points satisfying (3.12) are then the only minima of 
the potential, modulo gauge transformations in the $U(1)$ bundle.
The question of whether there exist maps $X$ and $Y$ for which 
a minimal connection (3.12) can be constructed as 

\begin{equation}
A_{0}=XdY.	
\end{equation}
still remains to be answered.

Furthermore, we have to determine how many maps yield a minimal 
connection.  It may happen that there are infinite independent maps, 
all of them giving rise to the same minimal connection, that is the 
old problem of valleys extending to infinite.
Before discussing these points we will completely characterize the minimal 
connections.  

\setcounter{equation}{0}\section{Minimal Connections: Monopoles over Riemann Surfaces}

We will show in this section that the minimal connections are the 
Dirac-Hopf monopole connections on $S^{2}$ and its generalization 
to topologically non-trivial Riemann surfaces found in \cite{M+R}\cite{F}.
We briefly discuss here their construction, for a more extensive 
analysis see \cite{M+R}\cite{F}.

The explicit expression of the monopole connections is obtained in terms of the 
abelian differential $d\tilde{\Phi}$ of the third kind over the 
compact Riemann surface $\Sigma$ of genus g.  $d\tilde{\Phi}$ is a 
meromorphic 1-form with poles of residue +1 and -1 at points $a$ and $b$, 
with real normalization. $\tilde{\Phi}$ is the abelian integral, 
its real part $G(z,\bar{z},a,b,t)$ is a harmonic univalent function 
over $\Sigma$ with logarithmic behavior around $a$ and $b$

\begin{eqnarray}
	\ln(\frac{1}{|z+ a|}) & +& \mbox{regular terms ,} \nonumber  \\
	\ln{|z-b|} & + & \mbox{regular terms ,} 
\end{eqnarray}

It is a conformally invariant geometrical object.  $z$ denotes the local 
coordinate over $\Sigma$ and $t$ the set of $3g-3$ parameters describing 
the moduli space of Riemann surfaces. We are considering maps from  
$\Sigma \mapsto S^{1}\times S^{1}\times M^{7}$ for a given $\Sigma$, so 
the parameters $t$ are kept fixed. They show however that the construction 
of minimal connections is a conformally invariant one.   

Let $a_{i}$, $i=1,...,m$ be m points over the compact Riemann 
surface.  We associate to them integer weights $\alpha_{i}$, 
$i=1,...,m$, satisfying

\begin{equation}
\sum_{i=1}^{m} \alpha _{i} =0
\end{equation}

We define
\begin{equation}
\phi = \sum_{i=1}^{m} \alpha _{i} G(z,\bar{z},a_{i},b,t).	
\end{equation}
and  have
\begin{eqnarray}
	\phi  & \rightarrow & - \infty \,\, \mbox{at $a_{i}$ with negative weights} 
	\nonumber  \\
	\phi  	 & \rightarrow & + \infty \,\, \mbox{ at $a_{i}$ with positive
	weights}.  	\nonumber 
\end{eqnarray}
$\alpha_{i}$ are integers in order to have univalent transition 
functions over the nontrivial fiber bundle that we consider.

We denote $\tilde{\Phi}$ the abelian integral with real part 
$\phi$.  Its imaginary part $\varphi$ is also harmonic but 
multivalued over $\Sigma$,
\begin{equation}
\tilde{\Phi}	= \phi +i\varphi.
\end{equation}

Let us consider the curve {\cal {C}} over $\Sigma$ defined by 

\begin{equation}
\phi =constant.\nonumber	
\end{equation}

It is a closed curve homologous to zero.  It divides the Riemann 
surface into two regions $U_{+}$ and $U_{-}$, where $U_{+}$ contains 
all the points $a_{i}$ with negative weights and $U_{-}$ the ones 
with positive weights.

We define over $U_{+}$ and $U_{-}$ the connection 1-forms
\begin{eqnarray}
	A_{+} & = & \frac{1}{2}(1+\tanh(\phi)) d\varphi \nonumber  \\
	A_{-} & = &\frac{1}{2}(-1+ \tanh(\phi))d\varphi
\end{eqnarray}
respectively.
$A_{+}$ is regular in $U_{+}$ and $A_{-}$ in $U_{-}$.
In the overlapping $U_{+} \bigcap U_{-}$ we have

\begin{equation}
A_{+}=A_{-}+d\varphi	
\end{equation}
$g= exp ({i\varphi})$ defines the transition function on the overlapping 
$U_{+} \bigcap U_{-}$, and because of the integer weights it is 
univalued over $U_{+} \bigcap U_{-}$. 

The base manifold $\Sigma$, the transition function $g$ and the 
structure group $U(1)$ have a unique class of equivalent $U(1)$ 
principle bundles over $\Sigma$ associated to them.  (4.6) defines a 
1-form connection over $\Sigma$ with curvature

\begin{equation}
F=\frac{1}{2}\frac{1}{\cosh{^2}{\phi}}d\varphi\wedge d\phi
\end{equation}

The $U(1)$ principle bundles are classified by the sum of the 
positive integer weights $\alpha _{i}$

\begin{equation}
N= \sum_{i}\alpha_{i}^{+}  \; \; , \alpha_{i}^{+} > 0.	
\end{equation}
which is the only integer determining the number of times $\varphi$ 
wraps around {\cal {C}}.  All the bundles with the same N are 
equivalent.
(4.8) satisfies (3.14).  Moreover they also satisfy (3.12).  In fact, 
since $\varphi$ and $\phi$ are harmonic over $\Sigma$, the metric is

\begin{equation}
d^{2}s=\frac{1}{\cosh ^{2} {\phi} }(d^{2}\varphi + d^{2}\phi),	
\end{equation}
and then (3.12) follows directly.

We have explicitly constructed all the $U(1)$ principle bundles over 
compact Riemann surfaces $\Sigma$, and  1-form connections 
(4.6) which describe strings of monopoles, generalizing the Dirac 
monopole construction.
The different connections obtained by choosing different sets of $a_{i}$, 
are all regular over $\Sigma$ and are different descriptions 
(using different metrics over $\Sigma$ ) of the same physical string 
of monopoles. All metrics are on the same conformal class, but with 
different number of points $a_i$, at which they become zero. This picture 
is a generalization of the construction of the Dirac-Hopf connection 
over $S^{2}$ in terms of the metric $d^{2}s=d^{2}\theta + sen^{2}
\theta d^{2}\varphi$ whose determinant becomes zero at the north and south 
pole. In the particular case of $\Sigma$ being $S^{2}$ with only one $a$ 
with weight $N$, (4.10) reduces to the standard $F=N sen\theta d\theta \wedge 
d\varphi$. 

\setcounter{equation}{0}\section{Reconstruction of Membrane  maps}

Given two minimal connections on the same principle bundle they 
differ by a closed 1-form, locally we have
\begin{equation}
A_{1}=A_{0}+d \Lambda .	
\end{equation}

We will consider now the residual area preserving diffeomorphisms 
left after the light cone gauge fixing, and show that we 
can locally generate any exact $d\Lambda$ by such a transformation, i.e the
residual symmetry on the membrane induces, at least locally, an U(1) gauge transformation 
on the space of connections.

The area preserving diffeomorphisms are generated by the first class 
constraints.
\begin{eqnarray}
	d(P_a dX_a) & = & 0 \\
	\oint_{C_{i}}P_a dX_a & = &0
\end{eqnarray}
where $P_a$ and $X_a$ are the transverse components of the membrane 
maps in the light cone coordinates.  $C_{i}$ is a basis of homology over the 
Riemann surface $\Sigma$.  In the right hand member of (5.3) we may 
add $2\pi n^{I}$ if we allow $X^{-}$ to take values on $S(1)$, in 
this case the homology must be integral. The infinitesimal gauge 
transformations generated by (5.2) and (5.3) may be expressed in terms of 
the Poisson bracket
\begin{equation}
\delta {F} =  \left\{F,\left\langle d\xi \wedge PdX\right\rangle_{\Sigma}¥
\right\} .\nonumber	
\end{equation}
where $\xi$ is the infinitesimal parameter of the transformation. $d\xi$ 
is a closed 1-form, so $\xi$ may be multivalued over $\Sigma$. The point 
is that we may reexpress 
\begin{equation}
\left\langle d\xi \wedge PdX\right\rangle_{\Sigma} =  \left\langle -\xi 
 dP\wedge dX\right\rangle_{\Sigma} +  \left\langle d (\xi 
 P dX\right\rangle_{\Sigma},\nonumber
\end{equation}
the first term in  the right hand side is the gauge transformation 
generated by (5.2), while the second term may be rewritten in terms of the 
transition of $\xi$ only, and represents then the gauge transformation 
generated by the global constraint (5.3). We will restrict  $d\xi$ to have 
integer periods in order to preserve the bundle structure we have 
introduced in previous sections. We will only use the gauge 
transformations generated by the local constraint.

We define over $U^{+}$ and $U^{-}$ on the membrane, $X_{0}$ and $Y_{0}$,
a pair of maps into the compactified directions of the target manifold, with

\begin{eqnarray}
X_{0}^{+}	 & = &  \frac{1}{2}(\tanh(\phi) + 1) \nonumber \\
X_{0}^{-}	& = & \frac{1}{2}(\tanh(\phi) -1)
\end{eqnarray}
respectively, and

\begin{equation}
dY_{0}=d \varphi ,	
\end{equation} 
where $\phi$ and  $\varphi$  are the two harmonic functions introduced in section
4. $X_{0}$ may then be expressed as a sum of a harmonic function

\begin{equation}
\frac{1}{2}(\phi \pm 1) 
\end{equation}
with a jump at  $U^{+} \bigcap U^{-}$, giving rise to a harmonic differential,
plus functions yielding an exact 1-form. We will call the pair $(X_{0},Y_{0})$  a
minimal map.
The minimal connection $\widehat{A}$ may then be expressed in terms of minimal
maps as 

\begin{equation}
\widehat{A} =X_{0}dY_{0} .	
\end{equation}

Let us now consider the change of $\widehat{A}$ under area preserving
diffeomorphisms. We obtain

\begin{equation}
\delta \widehat{A}_{r} = \partial_{r} (-\epsilon^{st}\partial_{t}\xi \widehat{A}_
{s} - \frac{1}{2}\xi {^\ast\widehat{F}} ),	 \end{equation}
where $\xi(\sigma^{1},\sigma^{2})$ is the infinitesimal parameter of the
transformation. (5.8) is only valid for a minimal connection $\widehat{A}$, 
since ${^\ast\widehat{F}}$ is then constant and may be included in the 
total derivative.  Given $\xi$, a gauge transformation, induced on the connection
space, is defined by

\begin{eqnarray}
\Lambda^{+}	 & = & -\epsilon^{st}\partial_{t}\xi \widehat{A}_
{s}^{+} - \frac{1}{2}\xi {^\ast\widehat{F}}  \\
\Lambda^{-}	& = & -\epsilon^{st}\partial_{t}\xi \widehat{A}_
{s}^{-} - \frac{1}{2}\xi {^\ast\widehat{F}} 
\end{eqnarray}
at $U^{+}$ and $ U^{-}$ respectively.

Conversely, given $\Lambda^{+}$ and $\Lambda^{-}$ defined on $U^{+}$ and $
U^{-}$ respectively, there exists a differentiable function $\xi$ satisfying 
(5.11) and (5.12). In fact, let us denote $C$ a closed curve over $U^{+} \bigcap U^{-}$
dividing $\Sigma$. This curve may be taken to be $\phi = 0$ without a lost of
generality. On $C$, we take the boundary condition

\begin{equation}
\xi \mid_{C}= - \frac{(\Lambda^{+} + \Lambda^{-})}{^\ast\widehat{F}} ,
\end{equation}
$^\ast\widehat{F}$ is different from zero because of the irreducibility condition.
We then solve (5.11) with boundary condition (5.13) on $U^{+}$ and (5.12) with
boundary condition (5.13) on $U^{-}$. The solution $\xi(\sigma^{1},\sigma^{2})$ is
then differentiable with continuous derivatives on $C$.

We thus conclude that the space of all minimal connections may be generated  by
considering the minimal connection (5.9) in terms of the minimal maps plus 
a representative of each real cohomology class of 1-forms over $\Sigma$.

We will now show that the space of maps which give rise to minimal 
connections is finite dimensional.

The general map $(X,Y)$ leading to a minimal connection $A$ may be written in terms of
a minimal map $(\widehat{X},\widehat{Y})$ as

\begin{eqnarray}
X	 & = & \widehat{X} + \tilde{x}      \nonumber \\
Y	& = & \widehat{Y} + \tilde{y}       ,
\end{eqnarray}
and the corresponding connection as

\begin{equation}
A= \widehat{A} + (\widehat{X}d\tilde{y} + \tilde{x}d\widehat{Y} + \tilde{x}d
\tilde{y} ) \end{equation}

For $A$ to be minimal, the parenthesis in (5.15) must be a closed 1-form. Hence
the pair $(\tilde{x},\tilde{y})$ is restricted by that condition. Let us 
assume that the term in the parenthesis in (5.15) is a closed 1-form. $\tilde{x}$
must then be an univalued map. If we keep $\tilde{y}$ fixed and consider 
$\tilde{x_1}$ a map giving rise to closed 1-form $(\widehat{X}d\tilde{y} +
 \tilde{x_1}d\widehat{Y} + \tilde{x_1}d\tilde{y} )$in the same cohomology 
 class then
 \begin{equation}
(\tilde{x} - \tilde{x_1})(d\widehat{Y} + d\tilde{y}) ,
\end{equation}
must be exact, and hence it may be annihilated by an area preserving 
diffeomorphism as shown previously. We may then ask how many $\tilde{x_1}$ 
are solutions of 
\begin{equation}
(\tilde{x} - \tilde{x_1})(d\widehat{Y} + d\tilde{y}) = 0 .
\end{equation}

The only solution not violating the irreducibility condition is 
$\tilde{x} = \tilde{x_1}$. The same argument may be performed leaving 
$\tilde{x}$ fixed and varying $\tilde{y}$.

We then come to the conclusion that the space of maps giving rise to 
a minimal connection is a finite dimensional space related to the space of 
cohomology classes of 1-forms. This means that there are no valleys at the 
minima since the latter correspond to an infinite dimensional space. We 
have not exhausted the residual gauge freedom related to the global 
constraint (5.3), we have only used the gauge transformation generated by 
the local constraint. The global constraint with the restriction we have 
assumed generates large gauge transformations, it transforms connections 
by adding elements of the integer cohomology class. Hence the space of 
minima may be further reduced to $H^{1}(\Sigma, R)/H^{1}(\Sigma, Z) $.

\setcounter{equation}{0}\section{Conclusions}

We showed that the Hamiltonian of the $D=11$ membrane satisfying the 
irreducibility condition, introduced in section 3, has isolated minima.
The space of these minima are classified by integer numbers 
defined by the irreducibility condition. For each $N$ there is an 
associated $U(1)$ bundle over the 2-dimensional worldvolume $\Sigma$ of 
arbitrary genus, and the minima are the monopole connections over Riemann 
surfaces \cite{M+R}\cite{F} . We explicitly constructed the minimal connections in 
terms of the membrane maps from $\Sigma \mapsto S^{1} \times S^{1}\times M^{7}$.

The minimal maps are strictly defined over a punctured compact Riemann surface, 
however the minimal connections are regular over the surface. The 
Hamiltonian density is consequently regular on it as well. We also showed , using area preserving 
diffeomorphisms, that there are no infinite dimensional valleys at the 
minima. In distinction to the Hamiltonian of the $D=11$ supermembrane with target 
space $M^{11}$ or $S^{1}\times M^{10}$ where there are infinite 
dimensional valleys at the minima giving rise to a continuous spectrum.

Outside the minima of the potential, it is possible to construct explicit configurations with $C^{\infty}$ maps 
satisfying the global irreducibility condition, and with a non uniform distribution of the 
winding density . Allowing for open neighborhoods on the 
membrane where there is no winding and where supersymmetric valley 
configurations could arise, i.e. solutions with thick hairs that go to 
infinity. The existence of these maps would then give  a continuous spectrum
as first argued in \cite{de Wit}. 

An interesting point to analyze now would be to completely characterize 
the potential of the irreducible membranes and supermembranes in a general 
way, since our study has been concerned only with the behavior of the 
potential near the minima. It will then be possible to answer the question 
of the continuity or discretness of the spectrum for the case of perhaps 
`non degenerate irreducible' supermembranes since our analysis shows that 
the global irreducibility condition is not enough to forbid the existence 
of valleys outside the minima, but it shows anyhow that it is a good 
condition to prohibit the valleys at the minima. Also,it points to the requirement of 
a non degeneracy condition on the 2-form integrated in the irreducibility 
condition. This 2-form being closed and non degenerate would define a symplectic 
geometry on the space of maps with winding.   
 
We would like to relate in general the structure of minimal maps and connections
to the central charge analysis of the SUSY algebra recently discussed in \cite{de Wit}.
In fact, it is easy to see that our global irreducibility condition is directly 
related to the central charges of the SUSY algebra. Also, the minimal maps 
and corresponding minimal connections satisfy the BPS condition with 
$\frac{1}{2}$ SUSY as in \cite{Ezawa} proving that they are supersymmetric
BPS states.

{\it Acknowledgements}

I. Martin likes to thank B. de Wit and J. Plefka for helpful conversations.


\begin{thebibliography}{99}

	\bibitem{N + de Wit} B. de Wit, M. L$\ddot{u}$scher and H. Nicolai, Nucl.
	Phys.{\bf B320 }(1989) 135	
	\bibitem{Russo} J.G. Russo, Nucl. Phys.{\bf B492 }(1997) 205,
	R. Kallosh, hep-th/9612004
	
	\bibitem{Pol} J. Polchinski, S. Chaudhuri and C.V. Johnson, hep-th/9602052;
	J. Polchinski, hep-th/9611050.
	\bibitem{de Wit} B. de Wit, K. Peeters and J. Plefka,  hep-th/9705225.
	
	\bibitem{C+M+R} M. Caicedo, I. Martin and A. Restuccia, hep-th/9701010.
	
	
	\bibitem{H} A. Weil,{\it Vari\'{e}t\'{e}s Kaehl\'{e}riennes}, Hermann (1957).
	 	
	\bibitem{M+R} I. Martin and A. Restuccia, Lett. Math. Phys.{\bf 39 }(4)
	(1997).
	\bibitem{F} F. Ferrari, hep-th/9310024.
	\bibitem{Ezawa} K. Ezawa, Y. Matsuo and K. Murakami,  hep-th/9706002. 
	
	
	
	

	
\end{thebibliography}
\end{document}